\documentclass[preprint, prb]{revtex4}

\usepackage{amsmath}
\newcommand{\third}{\mbox{\small{$\frac{1}{3}$}}}
\begin{document}

\title{Inconsistencies in the MIT bag model of hadrons}
\author{B. H. Lavenda}
\email{bernard.lavenda@unicam.it}
\affiliation{Universit$\grave{a}$ degli Studi, Camerino 62032 (MC) Italy}

\newcommand{\fourthirds}{\mbox{\small{$\frac{4}{3}$}}}
\newcommand{\fourth}{\mbox{\small{$\frac{1}{4}$}}}

\begin{abstract}
It is shown that what is commonly referred to as the MIT \lq bag\rq\ model of hadrons is thermodynamically wrong: The adiabatic conditions between pressure and temperature, and between pressure and volume imply the third, an adiabatic relation between temperature and volume. Consequently, the bag model is destitute of any predictive power since it reduces to a single adiabatic state.  The virial theorems proposed by the MIT group are shown to be the result of the normal power density of states of a non-degenerate gas and not the exponential density of states of the Hagedorn mass spectrum. A number of other elementary misconceptions and inaccuracies are also pointed out.

\vspace{25pt}
\end{abstract}
\maketitle

The MIT bag model~\cite{MIT} assumes that there is a region of space which contains hadronic fields which are acted upon by a constant, positive potential energy, $B$, per unit volume. This region of space is referred to as a \lq bag\rq. The constant $B$ of the bag is the maximum external pressure corresponding to the maximum temperature of the bag, or what is usually referred to as the  \lq Hagedorn\rq\ temperature.

Accepting these statements at face value, it follows that the internal energy of the gas and the volume which it occupies are at the same limit. Hence, the analogy with a first order phase transition in which the heat communicated to the system to change its volume, at constant temperature, so as to leave its vapor pressure unchanged is unfounded, because there can be no change in the volume in that limit. Using the method of characteristics we will give a proof that two adiabatic conditions implies the third. Moreover, we give a complete characterization of the photon gas on an adiabat which is not possible on an isotherm when the independent variables are chosen to be temperature and pressure. It is quite remarkable that the three infinite terms in the calorimetric equation, when pressure and temperature are chosen as independent variables,  involving the heat capacity at constant pressure,  the isothermal compressibility, and the coefficient of thermal expansion, exactly cancel one another on an adiabat. Finally, we show that their virial expressions  follow from a normal power law density of states and not the exponential Hagedorn mass spectrum, which they claim incorrectly to have derived from the thermodynamic formalism.

What the authors refer to as the total energy, $E$, in the first law
\begin{equation}
E=E_r+BV=4BV=H, \label{eq:I}
\end{equation}
is actually the enthalpy.  $E_r$ is the \lq radiation\rq, or internal, energy of the gas,
\begin{equation}
E_r=\sigma T^4V, \label{eq:Stefan}
\end{equation}
where $\sigma$ is the radiation constant, appears as the usual Stefan relation, but isn't because the temperature is held fixed at variable volume. We will now show that a consequence is that, in fact, the volume is also constant.

According to the second law, the increase in heat, as a function of enthalpy and pressure, is
\begin{equation}
dQ=T\,dS=dH-V\,dp. \label{eq:dQ}
\end{equation}
Due to the two phase nature of photons, the pressure is a function of temperature independent of the volume except on the adiabats. This is reflected in the fact certain thermodynamic quantities are infinite. If both temperature and pressure are kept constant, as they are in the bag model, $Q=H=\mbox{const.}$ Adiabaticity must be remain true for whatever independent variables are chosen.  Consequently, their second law 
\begin{equation}
dS=\frac{dQ}{T}=\fourthirds\sigma d\left(VT^3\right)=0, \label{eq:II}
\end{equation}
which in terms of their constant pressure $B$ states that
\begin{equation}
B^{3/4}V_0=\mbox{const.}, \label{eq:Vo}
\end{equation}
where $V_0$ is supposedly the \lq\lq volume containing one quantum particle.\rq\rq\
Rather, the volume has been shown to be constant, and equal to $V_0$. This is the only volume that the \lq bag\rq\ knows. 

We can reverse the argument: If (\ref{eq:Vo}) holds, and since $BT^{-4}=\mbox{const}.$,  it follows that
\begin{equation}
TV^{1/3}=\mbox{const}., \label{eq:ad}
\end{equation}
where
\begin{equation} 
T=T_0=\left(\frac{3B}{\sigma}\right)^{1/4}, \label{eq:To}
\end{equation}
and $V=V_0$. As Clerk Maxwell rightly asserted~\cite{Maxwell}: The thermal properties of a substance can only be defined completely when it is specified on both isotherms and adiabats. It is the adiabat (\ref{eq:ad}) that destroys the bag model.

We now offer the general proof that any two adiabatic relations implies the third. If $T$ and $p$ are chosen as independent variables in  (\ref{eq:dQ}), comparing coefficients of $dp$ on both sides of the equation gives
\[T\left(\frac{\partial S}{\partial p}\right)_T=\left(\frac{\partial H}{\partial p}\right)_T-V.\]
Introduce the Maxwell relation, $-(\partial V/\partial T)_p=(\partial S/\partial p)_T$, and use $H=4pV$, the third equality in (\ref{eq:I}). I then obtain
\begin{equation}
T\left(\frac{\partial V}{\partial T}\right)_p+4p\left(\frac{\partial V}{\partial p}\right)_T=-3V. \label{eq:pde}
\end{equation}

This partial de can be solved by the method of characteristics. The concept behind this is to identify paths in the $(T,p)$ plane, called characteristics, along which (\ref{eq:pde}) reduces to an ordinary de, which is much easier to solve. The auxiliary system of ordinary de, derived by Lagrange, is
\begin{equation}
\frac{dp}{4p}=\frac{dT}{T}=-\frac{dV}{3V}. \label{eq:Lagrange}
\end{equation}
Then the general solution to (\ref{eq:pde}) is
\begin{equation}\Phi(u,v)=0,\label{eq:gs}
\end{equation}
where $\Phi$ is an arbitrary function, and 
\begin{equation}
u=p^{1/4}/T=a \hspace{15pt}\mbox{and}\hspace{15pt} v=p^{3/4}V=b \label{eq:char}
\end{equation}
 are two independent solutions to (\ref{eq:Lagrange}), and $a$ and $b$ are arbitrary constants. At least one of these two independent solutions must contain the dependent variable, $V$. The remaining relation, $TV^{1/3}=c$, which is (\ref{eq:ad}), is obtained from (\ref{eq:char}) by eliminating $p$ between them, or directly by integrating (\ref{eq:Lagrange}).
 
 Alternatively, the general solution (\ref{eq:gs}) may be written as
 \begin{equation}
 p=T^4\phi\left(T^3V\right), \label{eq:gs-bis}
 \end{equation}
 where $\phi$ is another arbitrary function. The equation of state spans the entire range, from a photon gas, where $\phi=\mbox{const.}$, to an ideal gas where it is the inverse function.
 
The characteristics (\ref{eq:char}) are none other than adiabats defined by $dQ=0$. We begin with the fundamental calorimetric equation
\begin{equation}
dQ=L\,dV+C_v\,dT, \label{eq:cal}
\end{equation}
where $L=\varepsilon+p$ is the latent heat, $\varepsilon$ being the energy density, and $C_v$ is the heat capacity at constant volume. Transforming from the independent variables $V,T$ to $p,T$, I get
\begin{eqnarray}
dQ & =& -LV\kappa_T\,dp+C_p\,dT\nonumber\\
& = & -LV\left(\kappa_S+\frac{\alpha^2VT}{C_p}\right)\,dp+\left(C_v+\frac{\alpha^2VT}{\kappa_T}\right)\,dT \label{eq:cal-bis}
\end{eqnarray}
where $\kappa_{T,S}=-(1/V)(\partial V/\partial p)_{T,S}$ are the isothermal and adiabatic compressibilities, respectively, and $\alpha=(1/V)(\partial V/\partial T)_P$ is the coefficient of thermal expansion. Expression (\ref{eq:cal-bis}) contains three infinite terms, $\alpha$, $\kappa_T$ and $C_p$. However, on an adiabat, a photon gas is not infinitely compressible. It is remarkable that the infinities cancel along an adiabat.

The derivative of the pressure with respect to the temperature at constant volume and constant entropy are
\begin{subequations}
\begin{align}
\left(\frac{\partial p}{\partial T}\right)_V=\left(\frac{\partial p}{\partial T}\right)_S-\frac{C_v}{\alpha VT} \label{eq:pt-v}\\
\left(\frac{\partial p}{\partial T}\right)_S=\left(\frac{\partial p}{\partial T}\right)_V+\frac{3}{\kappa_T T}. \label{eq:pt-s}
\end{align}
\end{subequations}
Introducing the latter into the former gives
\begin{equation}
\frac{\alpha}{\kappa_T}=\frac{C_v}{3V}\hspace{25pt}\mbox{and}\hspace{25pt}
\frac{\alpha}{C_p}=\frac{\kappa_S}{3V}, \label{eq:con}
\end{equation}
where the second equality follows from the well-known relation $\kappa_T/C_p=\kappa_S/C_v$.
Because $\alpha$ and $\kappa_T$ are infinite for a photon gas mean that the partial derivatives in (\ref{eq:pt-v}) and (\ref{eq:pt-s}) are equal,
as can be seen by introducing (\ref{eq:con}) into (\ref{eq:cal-bis}). Setting it equal to zero I obtain
\begin{equation}
\left(\frac{\partial p}{\partial T}\right)_S=\frac{C_v(1+\third\alpha T)}{LV\kappa_S(1+\third\alpha T)}=\frac{C_v}{LV\kappa_S}=4\frac{p}{T}=\left(\frac{\partial p}{\partial T}\right)_V, \label{eq:Clap-Q}
\end{equation}
showing that the infinities in (\ref{eq:cal-bis}) cancel one another along an adiabat which is also an isochore, because the pressure is independent of the volume, and the adiabatic condition between pressure and temperature given in (\ref{eq:char}). 
However, this does not mean that $S=\mbox{const.}$ is equivalent to $V=\mbox{const.}$, as the bag model implies because the temperature is held constant. The adiabatic condition between temperature and pressure is given by (\ref{eq:ad}), and any two adiabatic conditions implies the third. The two adiabatic conditions given by the bag model, Eqn (2.8) or (\ref{eq:To}), and the equation following (2.8), equivalent to (\ref{eq:Vo}), imply $T_0^3V_0=\mbox{const.}$, and there is no other volume $V$.

Because the pressure is fixed, independent of variations in the volume, the analogy with a first order phase transition is illusory. When two phases are present, changes in the volume at constant temperature will make the liquid condense or evaporate so as to leave the vapor pressure unchanged. The Carnot-Clapeyron equations are~\cite{Lav05}
\begin{equation}
\left(\frac{\partial p}{\partial T}\right)_S=\frac{L}{T}=\frac{sC_v}{V}, \label{eq:CC}
\end{equation}
where $s$ is the Gr\"{u}neisen parameter, which for a photon gas equals $\third$.
Introducing (\ref{eq:CC}) into the fundamental calorimetric equation, (\ref{eq:cal}), yields
\[dQ=C_vT\,d\ln\left(V^sT\right)=0,\]
and no heat is communicated to the photon gas on an adiabat as a result of (\ref{eq:ad}).

In the MIT bag model $B$ is a pure constant independent of the state of the system. Through the absorption of heat at constant temperature work can be performed; but, as we know from Carnot, two different temperatures are required. In the bag model there is only a single temperature, $T_0$, and neither work can be performed nor heat absorbed. Variations in the volume at constant pressure and temperature are therefore completely illusory.

Turning to their variational principle in which the entropy
\begin{eqnarray}
S(E_r,V) & = & \fourthirds E_r^{1/3}(\sigma V)^{1/4}\nonumber\\
& = & \fourthirds(E-BV)^{3/4}(\sigma V)^{1/4}+S_0 \label{eq:var}
\end{eqnarray}
is to be maximized with respect to $V$ keeping $E$ fixed, we note: (1) From the condition that the temperature is held constant, the first equation says that the entropy is linear in the volume. (2) $E$ is really the enthalpy $H$ in the second equation, whose differential is
\begin{equation}
dH=T\,dS+V\,dp, \label{eq:H}
\end{equation}
so that if $p=B$, the constancy of $H$ demands that the entropy be constant. Consequently, there is no variational principle.

The adiabatic conditions, (\ref{eq:ad}) and (\ref{eq:char}), are preserved even when the bag is in motion at a uniform velocity, $v=c\beta$. This is a consequence of the fact that, out of all the thermodynamic potentials, $H$ is the only first order homogeneous function of the FitzGerald contraction, $\gamma^{-1}=\sqrt{1-\beta^2}$, viz.,
\begin{equation}
H=\frac{dH}{d\gamma^{-1}}\gamma^{-1}, \label{eq:H-bis}
\end{equation}
where the total derivative can be used since the other independent variables, $S$ and $p$, are Lorentz invariant. According to (\ref{eq:H}), (\ref{eq:H-bis}) implies that $\gamma T$ and $\gamma V$ are also Lorentz invariant. Consequently, in a state of uniform motion the adiabatic conditions, (\ref{eq:ad}) and (\ref{eq:char}), become~\cite{Planck}:
\[
\gamma^{4/3}TV^{1/3}=\mbox{const.},\hspace{15pt} p^{1/4}/\gamma T=\mbox{const.}, \hspace{15pt}\mbox{and}\hspace{15pt} p^{3/4}\gamma V=\mbox{const.} \]

The temperature $T_0$ is not \lq\lq equivalent to the average kinetic energy of the partons\rq\rq~\cite{MIT} because the average number of partons varies as the cube of the temperature. It is well-known that thermal equations of state, where the energy is proportional to the power of the temperature, allow for additional degrees of freedom other than translational, e.g. creation and annihilation of particles.  Their equation for the average number of partons, (2.8),
\begin{equation}
N=\frac{E_r}{T_0}=3\frac{B}{T_0}V=V/V_0, \label{eq:N}
\end{equation}
where 
\[V_0=(3B)^{-3/4}\sigma^{1/4},\]
is the second adiabatic condition in (\ref{eq:char}),
gives the impression that the average number of partons can be varied merely by varying the volume $V$. Rather, the average number of partons, for a fixed volume, varies as the cube of the absolute temperature. 

Moreover, any pressure that is constant, independent of the variables necessary to specify the thermodynamic state, will lead to inconsistencies when these variables are varied in a change from one thermodynamic state to another. This remark also pertains to the Poincar\'e \lq hydrostatic\rq\rq\ pressure in which he needed to add on to the electrostatic mass, $m_{\rm es}$, in order to convert it into the electromagnetic mass $m_{\rm em}$, viz., $\fourthirds m_{\rm es}=m_{\rm em}$.

Their virial theorem, Eqn (2.12), is exactly that of a perfect nondegenerate gas, and has nothing whatsoever to do with a density of states given by 
\begin{equation}
\zeta=\zeta_0e^{S}=\zeta_0e^{H/T_0} \label{eq:Hagedorn}
\end{equation}
where $\zeta_0$ is a constant. The density of states (\ref{eq:Hagedorn}) is simply a consequence of the fact that the entropy of degenerate systems does not have the logarithmic form  that a normal, power density of states has. Rather, (\ref{eq:Hagedorn}) is an asymptotic limit in which Cocconi's assumption, that both the internal energy and particle number diverge in such a way that their ratio remains finite, is satisfied.~\cite{Lav07}

We may rewrite their Eqn (2.12) in terms of kinetic theory as
\begin{equation}
B=\third mc^2\frac{N}{V}\left<\frac{\beta^2}{\sqrt{1-\beta^2}}\right>. \label{eq:B}
\end{equation}
 Now, introducing their relation (2.8), or my (\ref{eq:N}), into (\ref{eq:B}) results in
\[
\frac{T_0}{mc^2}=\left<\frac{\beta^2}{\sqrt{1-\beta^2}}\right>, \]
where their definition of the maximum temperature, (\ref{eq:To}), has been used.
Evaluating the average using the Maxwell-Juttner distribution~\cite{Juttner} for a nondegenerate perfect gas for which the number of particles in the range $\theta$ to $\theta+d\theta$ is
\begin{equation}
N(\theta)\,d\theta=\frac{V}{\pi^2\lambda^3_C}e^{-(mc^2/T_0)\cosh\theta}\sinh^2\theta\cosh\theta\,d\theta,\label{eq:MJ}
\end{equation}
where $\theta$ is the rapidity, defined by
\[\theta=\tanh^{-1}\beta,\]
and $\lambda_C$ is the Compton wavelength. Performing the average  gives $3BV_0=T_0$, or their Eqn (2.8). This clearly shows that the Maxwell-Juttner distribution (\ref{eq:MJ}) reproduces their results, and is completely extraneous to the Hagedorn mass spectrum (\ref{eq:Hagedorn}).

Even more can be said when we consider the average of the pressure
\begin{eqnarray}
B & = & \frac{mc^2}{\pi^2\lambda_C^3}\int_0^{\infty}e^{-(mc^2/T_0)\cosh\theta}\sinh^{4}\theta\,d\theta\nonumber\\
& = & \frac{mc^2}{\pi^2\lambda_C^3}\frac{K_2(x)}{x^2}, \label{eq:K}
\end{eqnarray}
where $K_2(x)$ is a Bessel function of order $2$, and $x=mc^2/T_0$. In the high and low  temperature limits
\begin{equation}
B\sim\frac{2}{\pi^2}\frac{T_0^4}{(\hbar c)^3} \label{eq:high}
\end{equation}
and
\begin{equation}
B\sim\frac{2T_0^{5/2}}{\lambda_C^3}\frac{e^{-mc^2/T_0}}{(2\pi mc^2)^{3/2}}, \label{eq:low}
\end{equation}
respectively. In the high temperature limit, (\ref{eq:high}) gives the pressure of an ultrarelativistic gas, while in the low temperature limit, (\ref{eq:low}) is the pressure of a degenerate boson gas. $B$ can hardly be considered as a constant external pressure when $T_0$ varies from the  nonrelativistic to the ultrarelativistic limits. Consequently, $T_0$ can hardly be considered as a limiting temperature.

\end{document}